\documentclass[aps,prc,twocolumn,showpacs,floatfix,nofootinbib,preprintnumbers,superscriptaddress,amsmath,amssymb]{revtex4-1}

\usepackage{epsfig}
\usepackage{color}
\usepackage{rotating}
\usepackage{natbib}
\usepackage{verbatim}
\usepackage{epstopdf}
\usepackage{graphicx}
\usepackage{changes}
\usepackage{bm}

\begin{document}

\title{A dynamical model calculation to reconcile the nuclear fission lifetime \\from different measurement techniques}

\author{M. T. Senthil Kannan}
\email{senthilthulasiram@gmail.com}
\affiliation{Department of Physics, Bharathiar University, Coimbatore-641046, India.}
\author{Jhilam Sadhukhan}
\email{jhilam@vecc.gov.in}
\affiliation{Physics Group, Variable Energy Cyclotron Center, Kolkata-700064, India.}
\affiliation{Homi Bhabha National Institute, Mumbai-400094, India.}
\author{B. K. Agrawal}
\affiliation{Saha Institute of Nuclear Physics, Kolkata-700064, India.}
\affiliation{Homi Bhabha National Institute, Mumbai-400094, India.}
\author{M. Balasubramaniam}
\affiliation{Department of Physics, Bharathiar University, Coimbatore-641046, India.}
\author{Santanu Pal}
\email{Formerly with Variable Energy Cyclotron Centre, Kolkata-700064, India.}
\affiliation{CS-6/1, Golf Green, Kolkata-700095, India.}
\date{\today}

\begin{abstract}

The pre-scission particle multiplicities suggest a lifetime of $\sim
10^{-20}$s for the  nuclear fission to occur which is in  contrast to
the fission lifetime $\sim 10^{-18}$s as predicted by  atomic probe.
This long standing ambiguity, arising due to the orders of magnitude
differences among  the fission lifetime measured from the nuclear
and atomic probes, has been addressed	within	a dynamical model which
includes the contributions from the nuclear shell effects.  We show that,
at lower excitation energies, these two probes decouples as the fissioning
system survives for a long time without any particle evaporation.
We also consider a wide range of  reactions to study the impact of
the excitation energy of compound nucleus on the fission dynamics in
general. Our model predicts the average fission life time of superheavy
nucleus $^{302}120$, to be more than $10^{-18}$s which is in reasonable
agreement with the recent experiments.

\end{abstract}

\maketitle


{\it Introduction} -- Nuclear fission is a fundamental decay mode for very
heavy atomic nuclei. The formation and survival of superheavy elements
\cite{bjo80,(kra12),oga99,oga15} is strongly governed by the associated
reaction dynamics and, in particular, the fission probability. Moreover,
the fission rate critically influences the origin of elements heavier
than iron \cite{mar07,gor13,sam18}. Therefore, a precise understanding
of the fission lifetimes is of extreme importance.

Substantial effort has been made to measure the
fission lifetime. Traditionally, the pre-scission
charged-particles \cite{les91,les93,kap17}, neutron
\cite{les93,kap17,hin92,hin86,hin89,hin93,siw95}, and $\gamma$-ray
\cite{pau94,van98,tho87,hof94} multiplicities are often used as a clock
to estimate the fission time. The transient time also can be estimated by
measuring the fission fragment distributions \cite{sch07}. In general, all
these nuclear techniques indicate that the fission process is fast enough
with an upper bound in average fission time: $\langle\tau_f\rangle\le
10^{-19}$s. These nuclear probe encompass different
variants of dissipative model to bridge the experimental observables
related to the fission lifetime. Often a simplified statistical model is
assumed to mimic the actual dynamics \cite{fro98,maz17}. On the other
hand, the blocking technique in single crystals, which is considered
to be a direct probe, leads to scission time scales much longer than
the ones inferred from the nuclear methods \cite{bro68,and76}. This
contradiction is intensified after several recent crystal-blocking
measurements \cite{mor98,and07,mor08,and08} that indicate
attosecond ($10^{-18}$s) time delay in heavy-ion induced fission. It
is shown recently that K X-ray emission prior to fission can be used
to measure fission lifetimes \cite{wil04,fre12}. This method measures
the long fission-time component in agreement with the crystal-blocking
technique. Both the atomic clocks are used to explore the survival of
superheavy element with $Z=120$ \cite{mor08,fre12}. A detailed review
on the study of fission lifetime can be found in Ref. \cite{jac09}.

 Theoretical modeling of fission is extremely challenging as it involves
 many-body quantum dynamics. Although time-dependent density functional
 theory methods may seem to be the most natural choice to describe
 this process, but, reconstructing entire distributions can become
 prohibitively expensive especially when pairing correlations are
 fully taken into account \cite{bul16}. Because each such calculation
 simulates only a single fission event. The situation becomes more
 complicated for induced fission from  excited states, where pairing is
 quenched and  dynamics becomes strongly dissipative and non-adiabatic
 \cite{nor83}. In this regime, stochastic transport theories have been
 successfully applied  to describe the energy transfer between the
 collective and intrinsic degrees of freedom of the fissioning nucleus
 \cite{abe96,fro98,sie86}. Among such theories, dynamical approaches based
 on the Langevin equation and its derivatives have been successful in
 reproducing fission dynamics \cite{nad12,ari14,ran15,den17,maz17,sie17}.

 In the present work, an explicit simulation of large amplitude collective
 dynamics is performed to extricate the long standing ambiguities of the
 fission lifetime extracted using the atomic and nuclear-technique
 measurements.


{\it Theoretical framework} -- 
We have implemented a state-of-the-art model based on  stochastic Langevin
equations to study the full dynamical evolution of an excited compound
system starting from the ground-state configuration up to the scission.
This approach allows one to account for the dynamical effects in
more realistic manner than those  in the combined dynamical and
and statistical model (CDSM) calculations \cite{fro98,laz93,gon02}.
Evaporation of light particles - $n$, $p$, $\alpha$ and statistical
$\gamma$-rays are sampled at each time step by using the Monte-Carlo
technique. The standard statistical model prescriptions for
particles \cite{wei37} and $\gamma$-ray \cite{(lyn68)}
evaporation are used for this purpose. The Helmholtz free energy $F$
is used as the driving force for the collective motion. 
Specifically, we assumed
the Fermi gas model \cite{(boh69)} to define: $F=V-(a-a_0)T^2$,
where $V$ and $a$ are the deformation dependent potential energy
and level density parameter \cite{rei81}, respectively, $a_0$
being the value of $a$ at the ground-state deformation. The energy
and deformation dependent shell-correction is incorporated in $a$
following Ignatyuk's prescription \cite{ign75}. The temperature
$T$ is obtained from the ground-state excitation energy $E^*$ as:
$T=\sqrt{E^*/a_0}$. The average liquid-drop part of $V$ is calculated
following the double-folding Yukawa-plus-exponential model \cite{sie86}
and the associated shell-correction energy is obtained by applying
the Strutinsky's method \cite{str68,bra72} of shell-correction to
the nucleonic levels generated with the two-centered Woods-Saxon
mean field~\cite{gar99}. We use the BCS pairing to account the nuclear
super-fluidity \cite{bra72,gar99}. Calculated potential barriers are found
to be in good agreement with the existing results \cite{dud84}. The
collective inertia for the Langevin dynamics is extracted using
Werner-Wheeler approximation \cite{dav77} for the irrotational flow of
incompressible nuclear fluid. The chaos-weighted wall friction model
\cite{pal98} is used to calculate the friction tensor as it seems to be
most suitable for the present purpose \cite{wad93,cha02,cha01}.

Large scale computing is performed for an ensemble of ~$10^6$ Langevin
trajectories to generate the results for a single macrostate. Each of
these trajectories is allowed a maximum dynamical time of $10^{-15}$s in
steps of $3 \times10^{-25}$s. In case of any particle evaporation, the
compound system and the associated dynamical quantities are modified
accordingly. Therefore, in advance, we calculate all the inputs
for a total of 48 daughter nuclei that leaves open the possibility
of fifteen neutron  ($n$) evaporations in combination with either a
proton ($p$) or an $\alpha$ evaporation. It ensures the scope of all
the feasible evaporation channels for the present study. For each
fission event, we record the average $n$-evaporation time $\tau_n$, the time $\tau_{nl}$ when the last $n$ is evaporated, and the
scission time $\tau_f$. According to the neutron-clock \cite{hin92},
the product: $\langle\tau_{n1}\rangle=n_{pre}\langle\tau_n\rangle$
gives the average fission lifetime, where $n_{pre}$ is the average
pre-scission neutron multiplicity. In practice, $n$-decay width,
which is directly related to $\tau_n$, is calculated and combined
with a suitable fission-decay model to reproduce the experimental
$n_{pre}$. Effectively, the measured $n_{pre}$ determines the experimental
fission lifetime \cite{les91,les93,hin92}. Equivalently, the $n$-clock
can be devised using $\tau_{nl}$ with the underlying assumption that
scission occurs immediately after the last neutron is evaporated. Of
course, the first-chance fission events are not sensitive to neutron
evaporations. We compare these two neutron probes with the actual
scission time $\tau_f$. In addition, for a comprehensive understanding
of the dynamical evolution, we calculate the average deformation $\langle d_f\rangle$
of the fissioning system by taking the time average of the collective
coordinate for each event.

\begin{figure}[!ht]
\includegraphics[width=1.0\columnwidth]{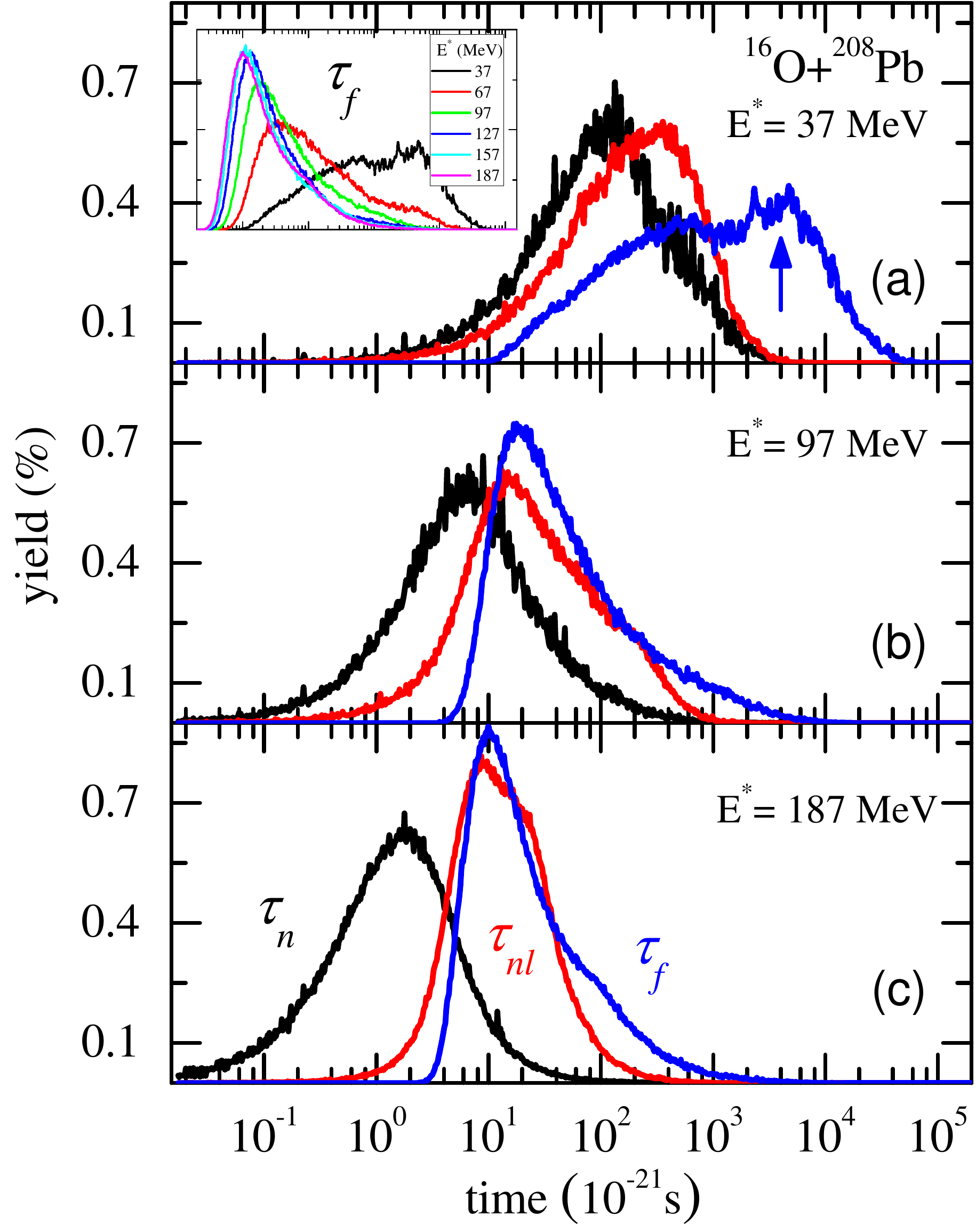}
\caption[C1]{\label{fig:EFP}
(Color online) Distributions of $\tau_f$, $\tau_{nl}$, and $\tau_n$
as labelled for initial excitation energy (a) $E^{*}=$ 37 MeV, (b)
$E^{*}=$ 97 MeV, and (c) $E^{*}=$ 187 MeV. 
Inset in the panel (a) depicts the changeover
of $\tau_f$-distributions with $E^*$. The peak of $\tau_f$-distribution
for $E^*=$37 MeV is indicated by arrow.} 
\end{figure}

\begin{figure}[!htb]
\includegraphics[width=1.0\columnwidth]{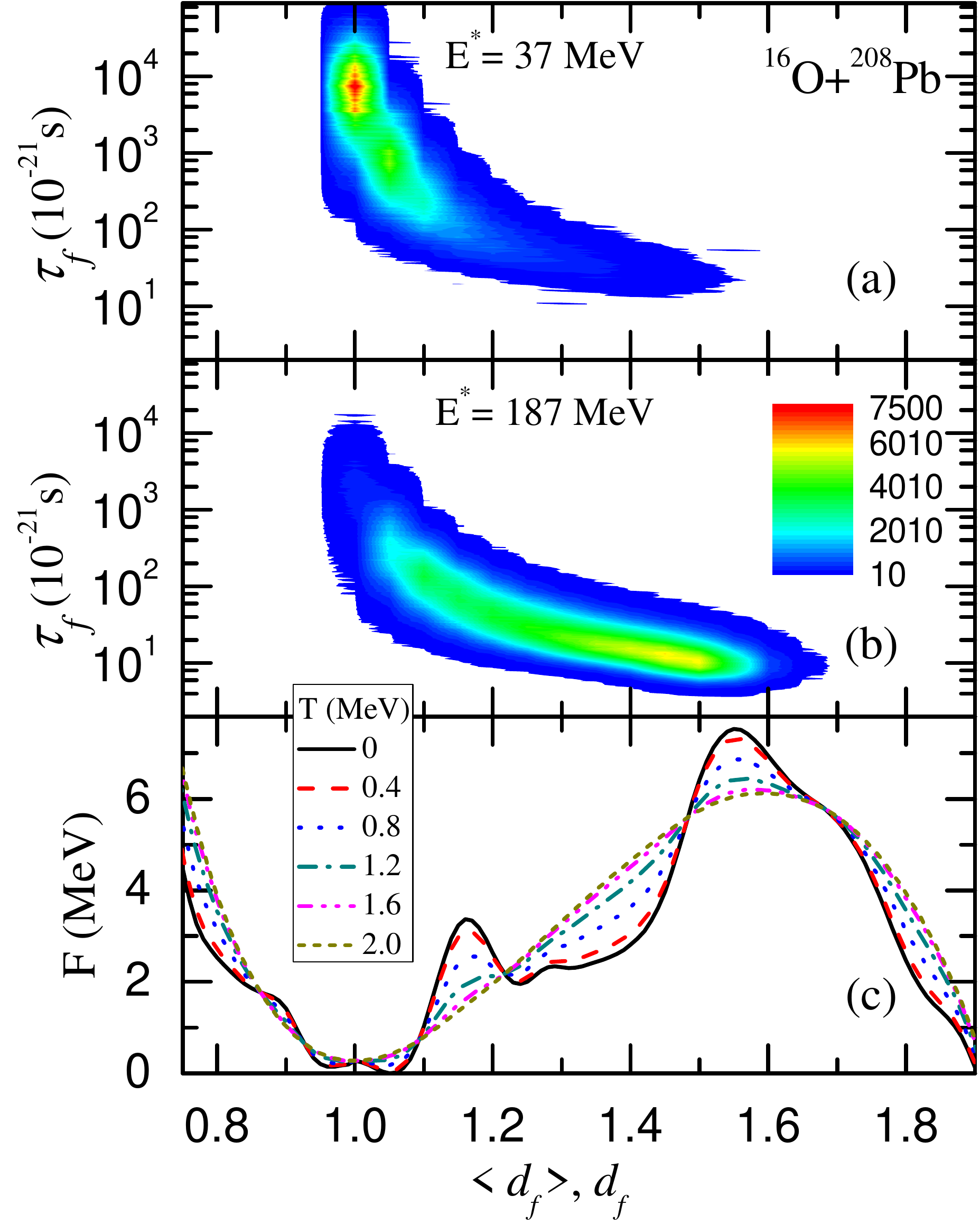}
\caption{(Color online) Distribution of fission yields on the $\tau_f-\langle d_f \rangle$
plane for (a) $E^{*}=$ 37 MeV and (b) $E^{*}=$ 187 MeV. (c) The variation
of free energy $F$ with deformation $d_f$ for different temperature as
indicated.} \end{figure}


{\it Results and Discussion} --  We first consider the
$^{16}$O$+^{208}$Pb$\rightarrow ^{224}$Th  reaction since this system is
well-studied experimentally \cite{hin92,ros92}. The normalized yields
corresponding to $\tau_n$, $\tau_{nl}$, and $\tau_f$ calculated
for different values of initial excitation energy $E^{*}$ are
shown in Fig. 1. Evidently, at large $E^{*}$ (Fig. 1(b)-(c)), the
distributions of $\tau_{nl}$ and $\tau_{f}$ almost coincide except
for the long-time tail in $\tau_f$. This behavior of $\tau_f$ is reported
in Refs. ~\cite{fro98,gon02} using CDSM calculations. In contrast, for
the lowest $E^*$ (Fig. 1(a)),  the shape of $\tau_f$ becomes very broad
with the peak at $\tau_f>10^{-18}$s, whereas the shapes of $\tau_{n}$
and $\tau_{nl}$ remain almost unaffected. This decoupling of $\tau_f$
from $\tau_{nl}$ and $\tau_n$ appears somewhere between $E^{*}=$ 67
MeV and 37 MeV (inset of Fig. 1(a)). It emphasizes the fact that, at
a lower energy, the fissioning system survives for a long time without
any particle evaporation as the available excitation energy falls below
the particle emission threshold. Additionally, we found that the long
fission-time component is further enhanced by the nuclear shell effects
as conjectured in \cite{fro98,gon02}.

\begin{figure}[tb]
\includegraphics[width=1.0\columnwidth]{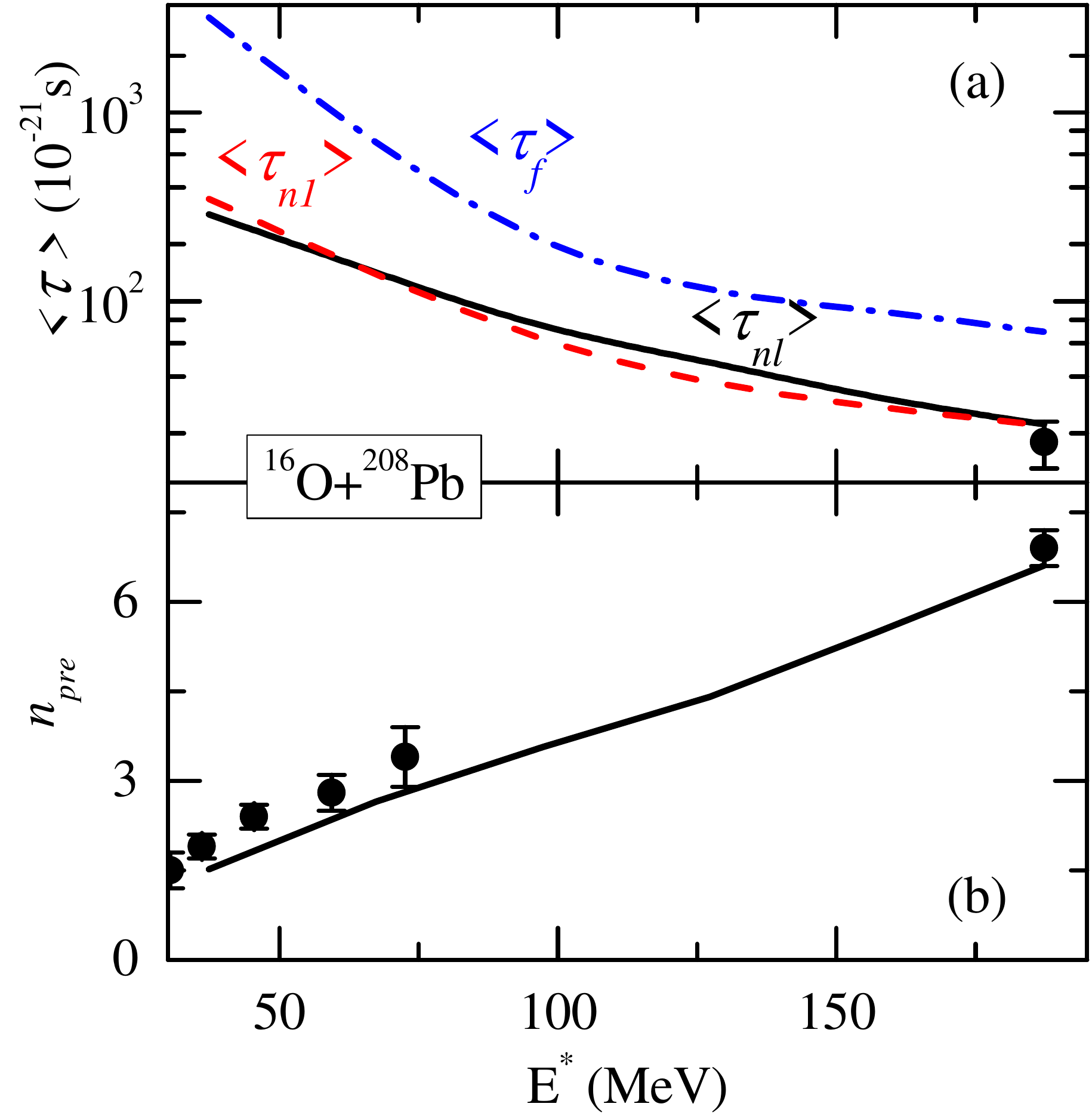}
\caption[C1]{\label{fig:FPY} (Color online) (a) Average fission lifetime
$\langle\tau_f\rangle$ (dash-dotted line), $\langle\tau_{n1}\rangle$
(dashed line), $\langle\tau_{nl}\rangle$ (solid line) as a function of
$E^*$. The symbol indicates the experimental data \cite{hin92}. (b)
Comparison of experimental \cite{ros92,hin92} $n_{pre}$ with the
calculated values.} \end{figure}


For a deeper understanding of the nature of $\tau_f$, we calculated the
correlation between $\tau_f$ and $<d_f>$. The corresponding two-dimensional
distribution of fission events are plotted in Fig. 2(a) and Fig. 2(b),
respectively, for the lowest and highest $E^{*}$ considered in
Fig. 1. Also, the free energy $F$ for different values of $T$ are shown
in Fig. 2(c). Clearly, the events with a long fission-time predominantly
stay around the ground state deformation ($0.95\le d_f\le 1.1$). Here,
$d_f=1$ corresponds to the spherical configuration. On the other hand,
the average deformation increases for the high energy fission events
as the free energy profile becomes flatter. This observation clarifies
the ambiguity related to the role of deformation-dependent dissipation
in escalating fission lifetime. Since majority of the long-time events
roam around the ground state deformation, these are hardly affected by
the dissipation near scission.

We have calculated the average fission
time $\langle\tau_f\rangle$, $\langle\tau_{nl}\rangle$, and
$\langle\tau_{n1}\rangle$ associated to the distributions of
$\tau_f$, $\tau_{nl}$, and $\tau_{n}$, respectively. These are
plotted in Fig. 3(a) along with $n_{pre}$ in Fig. 3(b). As expected,
$\langle\tau_{nl}\rangle$ and $\langle\tau_{n1}\rangle$ remain very close
to each other at all energies. For higher $E^*$, $\langle\tau_f\rangle$
is comparatively large due to the presence of long-time tail. Where as, 
at low $E^*$,  it is one order of magnitude higher than those
for the other two distributions due to the decoupling of $\tau_f$
from $\tau_{nl}$ and $\tau_n$ as  shown in   
Fig. 1(a). Moreover,
the absolute value of $\langle\tau_f\rangle$ reaches the attosecond
time-scale in agreement with atomic measurements. One experimental data
of fission lifetime is available for the present system from neutron
multiplicity measurement and it matches well with our calculated
$\langle\tau_{nl}\rangle$ and $\langle\tau_{n1}\rangle$. It is clear from Fig. 3(b), the experimental neutron multiplicities
are reproduced simultaneously without any free parameter.

\begin{figure}[!htb]
\includegraphics[width=1.0\columnwidth]{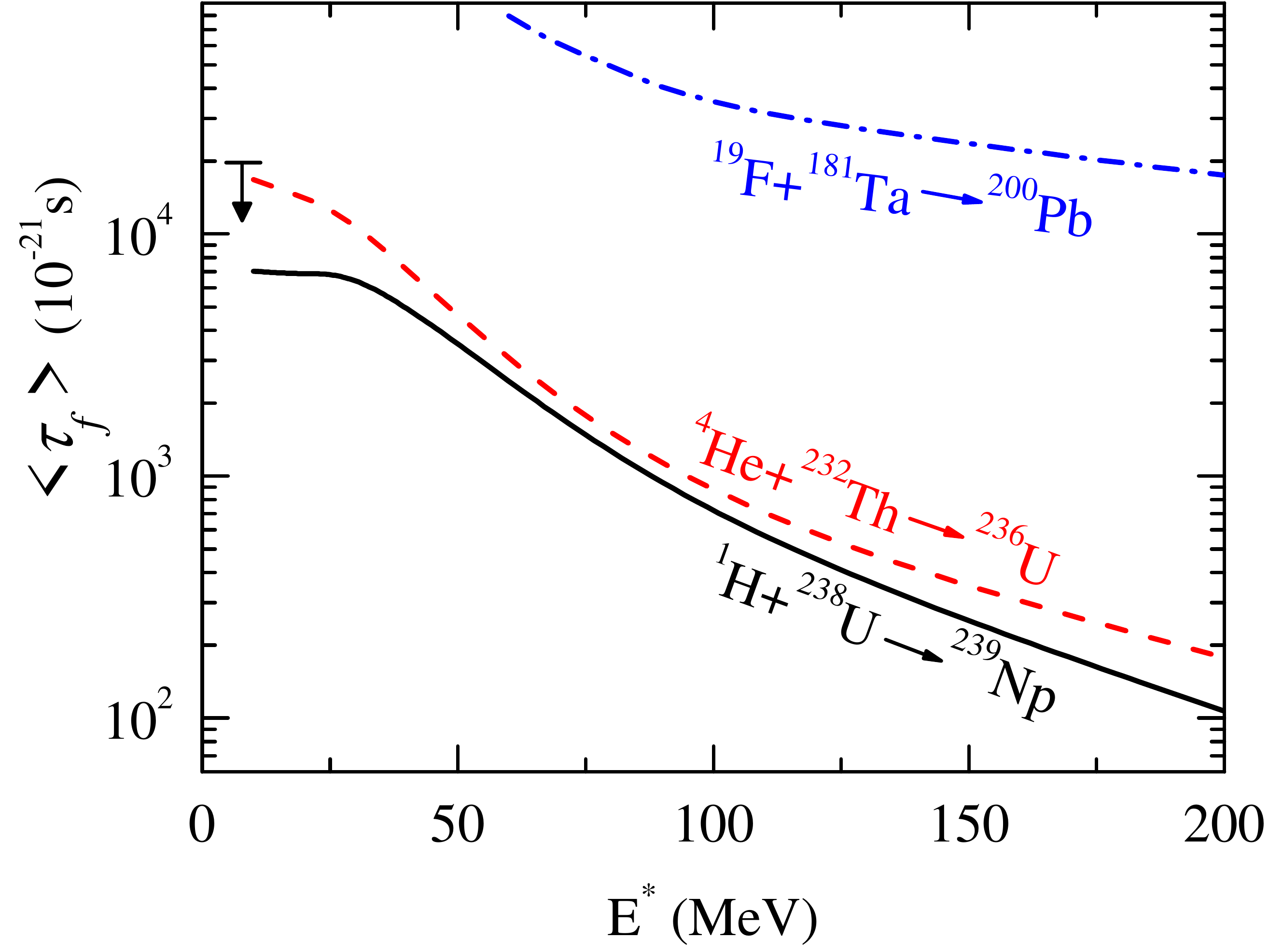}
\caption{(Color online) Average fission lifetime $\langle\tau_f\rangle$ as a function of excitation energy
for all the reactions mentioned in text.}
\end{figure}
 
After the benchmark study on $^{224}$Th, we computed the fission lifetime
for several other reactions. Specifically, we considered the reactions:
(1) $p+^{238}$U, (2) $^{4}$He+$^{232}$Th, and (3) $^{19}$F+$^{181}$Ta
covering a wide mass range relevant to fission. The excitation
energy dependence of $\langle\tau_f\rangle$ for these reactions are
demonstrated in Fig. 4. The $\langle\tau_f\rangle$ for both the
actinides vary similarly and become slower than a attosecond for
$E^*\le$ 90 MeV. This behavior is in compliance with the predictions
from the atomic probe. The upper limit of $\langle\tau_f\rangle$
for the reaction (1) is measured \cite{bro68} for the lowest $E^*$
(indicated by down arrow in Fig. 4). As evident in Fig. 4, our
calculation follows this limit. For reaction (3), $\langle\tau_f\rangle$
remains more than $10^{-17}$s even at a very large $E^*$. This system
has been studied extensively \cite{for87} around $E^*=$100 MeV and,
subsequently, analyzed theoretically in Ref. \cite{laz93}. In their
analysis, a comparatively large reduced friction was required within the
dynamical model to delay fission. However, our calculation reproduces
the lifetime of more than $10^{-17}$s without
recoursing to the tunning of any input parameter. To
determine the origin of large fission lifetime, we extracted the
distributions of $\tau_f$, $\tau_{nl}$, and $\tau_n$ (similar to Fig. 1)
in Fig. 5. Interestingly, $\tau_f$ has substantial contribution between
$10^{-17}-10^{-15}$s that results in a large $\langle\tau_f\rangle$. The
presence of this broad second peak is consistent with the crystal-blocking
data in Ref.~\cite{for87}.

\begin{figure}[!htb]
\includegraphics[width=1.0\columnwidth]{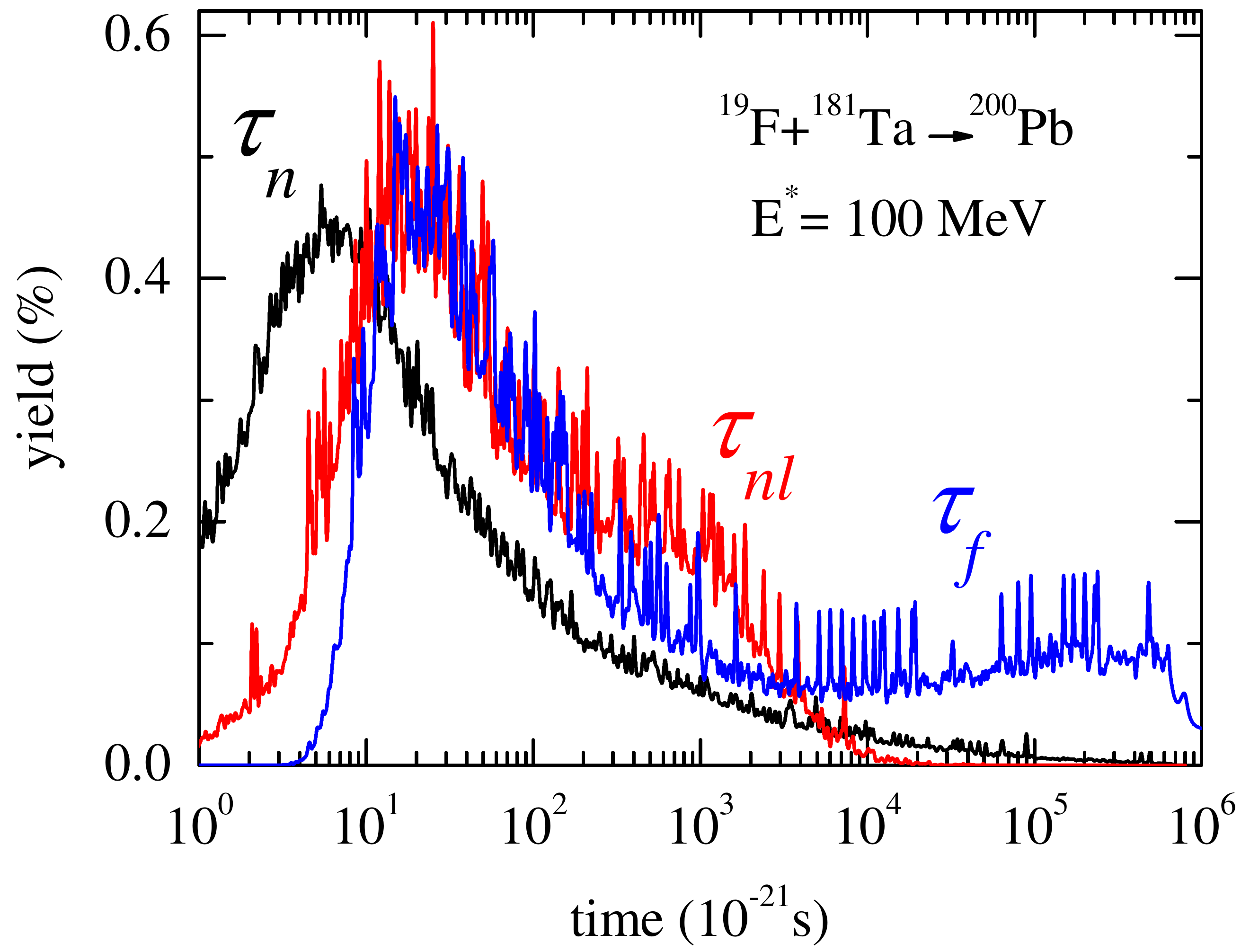}
\caption{(Color online) Distributions of $\tau_f$, $\tau_{nl}$,
and $\tau_n$ for the reaction $^{19}$F+$^{181}$Ta.} 
\end{figure}

Finally, we address the fission lifetime for the $^{238}$U+$^{64}$Ni
reaction which is proposed to be a possible candidate for the
discovery of $Z=120$ isotopes. Several studies have been made to this
end \cite{mor08,fre12,sik16}. Crystal blocking measurement \cite{mor08}
predicts that the $^{302}$120 nuclei can survive more than $10^{-18}$s. In
this experiment, the initial excitation energy of the compound system
was uncertain due to the large target thickness. We found, as shown
in Fig. 6, that the $\langle\tau_f\rangle\ge$ 10$^{-18}$s is possible
for this system only if the excitation energy $E^*\le 10$ MeV (dashed
lines). Therefore, the long-lived component in the experiment may be
contributed from very low energy events. The stability of superheavy
elements is very much sensitive to nuclear shell effects. Thus, an
accurate estimation of nuclear potential energy surface is essential for
an reliable prediction. In our calculation, the average fission barrier
for $Z=120$ isotopes is ~9 MeV which is close to the microscopic density
functional prediction \cite{bar15}. The multidimensional macro-micro
prediction \cite{mol09} is ~2 MeV smaller than this value. Although, the fission lifetime is strongly influenced by the fission barrier,
the temperature dependence of the potential surface apparently dilutes
this effect. So, a proper modeling of the finite-temperature potential is
crucial. Within the Fermi-gas model, the shell effects washes out much
faster, as shown in Fig. 2(c), compared to the finite-temperature shell
model predictions \cite{bra74}. To explore the effect of shell-washing
in the present system, we performed another set of calculation with a
modified shell damping factor.  Specifically, the energy dependence of the
level density parameter $a$ is reduced to 50$\%$ of its unaffected value.

The corresponding $\langle\tau_f\rangle$ is
plotted in Fig. 6  and it shows that, with this reduction, the attosecond
time can be reached at a higher $E^*\sim$ 30 MeV (solid line). This study
reveals the importance of an appropriate microscopic calculation as far
as superheavy elements are concerned.

\begin{figure}[!htb]
\includegraphics[width=1.0\columnwidth]{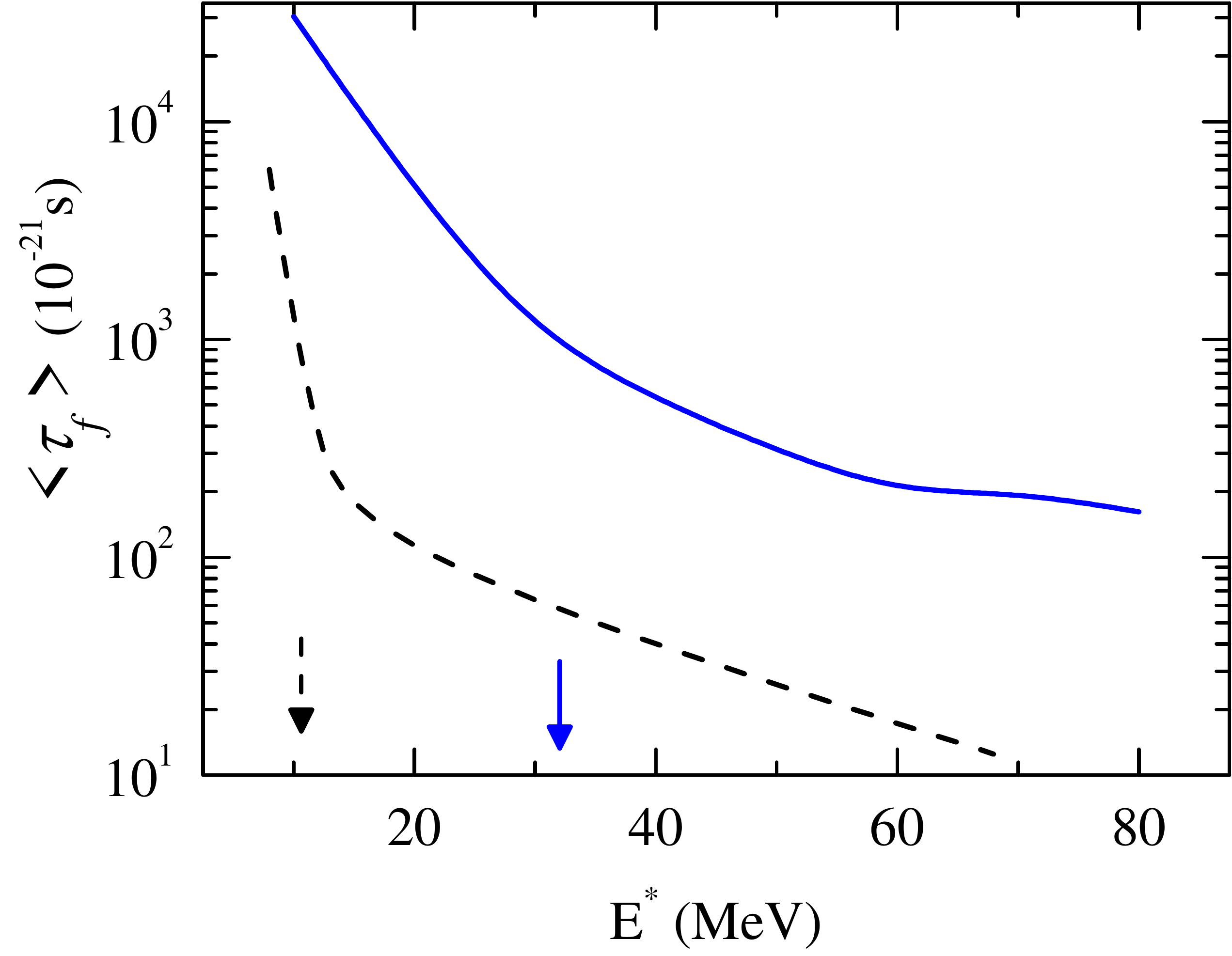}
\caption{(Color online)  Average fission lifetime as a function of excitation
energy calculated with full (dashed line) and reduced (solid line)
energy dependence in shell correction. Vertical arrows indicates the
$E^*$ corresponding to $\langle\tau_f\rangle=10^{-18}$s.} 
\end{figure}

{\it Conclusions} -- A state-of-the-art calculation of fission dynamics
is performed for excited compound systems where - (i) the dynamics
is followed from the ground state deformation to scission including
all possible evaporation channels, (ii) the energy and deformation dependent
shell-effect is properly accounted, and (iii) no parameters are tuned
to reproduce a specific observable. The average fission lifetime for a
wide range of reactions is calculated and found to be consistent with
atomic measurements. The neutron multiplicity as a probe for fission
lifetime is shown to be inaccurate for low excitation energies. Finally,
the attosecond lifetime for $Z=120$ nucleus is found to be consistent
with our theoretical calculations.

\begin{acknowledgments} 
Discussion with Nicolas Schunck is gratefully acknowledged. Computing
support for this work came from the Lawrence Livermore National Laboratory
(LLNL) Institutional Computing Grand Challenge program and the computing
facility at VECC. MTS acknowledges the support and warm hospitality from CNT/VECC during his stay at VECC.  \end{acknowledgments}

\bibliographystyle{apsrev4-1}
\bibliography{ref,books}
\end{document}